\documentclass[fleqn,twoside]{article}
\usepackage{espcrc2}

\usepackage{graphicx}
\usepackage[figuresright]{rotating}

\newcommand{\AmS}{{\protect\the\textfont2
  A\kern-.1667em\lower.5ex\hbox{M}\kern-.125emS}}

\title{The abelian projection revisited}

\author{A. Di Giacomo\address[MCSD]{Dipartimento di Fisica, Universit\`a di Pisa,\\
       I.N.F.N. Sezione di Pisa
       Via buonarroti 2, 56127 Pisa, Italy\\
        adriano.digiacomo@df.unipi.it, paffuti@df.unipi.it}%
        \thanks{Talk presented by A. Di Giacomo.},
        G. Paffuti\addressmark \\
        }
       
\begin{document}

\begin{abstract}
Abelian projection is reanalysed in the frame of the $SU(N)$ Higgs model.
The extension to QCD is discussed. It is shown that dual superconductivity
of the vacuum is an intrinsic property independent of the choice of the abelian projection.
\vspace{1pc}
\end{abstract}

\maketitle

\section{Introduction}
The mechanism of color confinement by dual superconductivity of the 
vacuum\cite{hooft75,mand75} requires the identification in QCD of a magnetic
$U_M(1)$ gauge field, which has to be a color singlet if magnetic charges have 
to condense in the vacuum and preserve the color gauge symmetry.

For $T < T_c$ $U_M(1)$ has to be Higgs broken (dual superconductivity),
for $T> T_c$ it must be restored (normal ground state). $U_M(1)$ is defined by a procedure known 
as ``Abelian Projection''\cite{hooft81}.

In sect.2 we  reanalyse the Abelian Projection in the $SU(N)$ Higgs model.
In sect.3 we  discuss the extension to QCD and we  show that dual
superconductivity (or not) is an intrinsic property independent of the 
particular choice of the projection.

\section{$SU(N)$ Higgs model.}
Consider the $SU(N)$ Higgs model. In the usual notation
\[ {\cal L} =  Tr\left\{
D_\mu \phi^\dagger D^\mu\phi\right\} - \frac{1}{4}
Tr\left\{
G_{\mu\nu} G^{\mu\nu}\right\} - V(\phi)\]
$\phi = \sum_{i=1}^{N^2-1} \phi_i T^i$ is a scalar field
in the adjoint representation. In the Higgs phase
$\langle \phi^a\rangle = \varphi^a \neq 0$, 
$\varphi^a$ being a nontrivial minimum of $V(\phi)$.

Consider the field strength tensor\cite{hooft74}
\begin{equation}
F^a_{\mu\nu} =
Tr\left\{ \phi^a G_{\mu\nu}\right\} - \frac{i}{g}
Tr\left\{ \phi^a \left[D_\mu\phi^a, D_\nu\phi^a\right]\right\}
\label{eq1}\end{equation}
$F^a_{\mu\nu}$ is a color singlet and gauge invariant, and such are
separately the two terms of eq.(1).
The following theorem has been proved in ref.\cite{deldesun}\\
\vskip0.2in
{\bf Theorem.}
Bilinear terms in $A_\mu A_\nu$ cancel between
the two terms
on the right side of eq(\ref{eq1}) and $F^a_{\mu\nu}$
obeys Bianchi identities ($\partial_\mu F^{a*}_{\mu\nu} = 0 $)
iff 
\begin{equation}
\phi^a = U(x) \phi^a_{diag} U^\dagger(x)
\label{eq2}\end{equation}
with $U(x)$ a generic gauge transformation and
\begin{equation}
\phi_{diag}^a = 
 diag\left(
\overbrace{\frac{N-a}{N},..,\frac{N-a}{N}}^{a},
\overbrace{-\frac{a}{N},..,-\frac{a}{N}}^{N-a}\right)
\label{eq3}\end{equation}
\vskip0.25in
The invariance group of $\phi_{diag}^a$ is $g =
SU(a)\otimes SU(N-a)\otimes U(1)$. It identifies a symmetric space\cite{weinb},
in the sense that if $L_0$ is the Lie algebra of the subgroup $g$, $L$ the algebra of $SU(N)$
and $L_1 = L-L_0$, then $\left[ L_0,L_0\right] \subset L_0$,
$\left[ L_0,L_1\right] \subset L_1$, $\left[ L_1,L_1\right] \subset L_0$.

Viceversa there is a conjecture by Michel\cite{michel,ling} that if the Higgs field belongs to the adjoint 
representation any $\phi^a$ identifies a symmetric space, or has the form eq.(\ref{eq2}),eq.(\ref{eq3}).

If $\phi^a$ is of the form of eq.(2), eq.(1) reduces identically to the form
\begin{eqnarray}
F^a_{\mu\nu} =
\partial_\mu Tr\left\{ \phi^a A_\nu\right\} - 
\partial_\nu Tr\left\{ \phi^a A_\mu\right\}\nonumber\\
-\frac{i}{g}
Tr\left\{ \phi^a \left[\partial_\mu\phi^a, \partial_\nu\phi^a\right]\right\}
\label{eq4}\end{eqnarray}
which is again gauge invariant and color singlet. In the gauge $\phi^a = \phi^a_{diag}$
(unitary gauge) $F^a_{\mu\nu}$ reduces to the abelian form
\begin{equation}
F^a_{\mu\nu} =
\partial_\mu  A^a_\nu - 
\partial_\nu  A^a_\mu
\label{eq5}\end{equation}
with
$A^a_\mu = Tr\left\{ \phi^a A_\mu\right\}$, whence the name ``abelian projection'' of the gauge transformation
to the unitary gauge.

$A^a_\mu$ is defined by expanding
diagonal part of $A_\mu$ in terms of roots
$A_{\mu , diag} = \sum_i \alpha^i A_\mu^i$
\begin{equation}
\alpha^i = diag(0,0,0\ldots\stackrel{i}{1},\stackrel{i+1}{-1},0\ldots 0)
\label{eq6}\end{equation}
$Tr\left\{ \phi^a_{diag} \alpha^b\right\} =\delta^{ab}$.

In the Higgs phase $\phi^a\neq0$ monopoles exist as solitons. The construction is the same
as in ref's\cite{hooft74,poly} for $SU(2)$, in the $SU(2)$ subgroup of $SU(N)$ spanned by the $i,i+1$
elements of eq.(\ref{eq6}). For these solutions the electric field $F^a_{0i} = 0$, the magnetic 
field at large distances $H^a_i = \frac{1}{2}\varepsilon_{ijk} F_{jk}$ is that of a Dirac monopole of charge $1/g$
\begin{equation}
\vec H = \frac{1}{g}\frac{\vec r}{r^3} + \hbox{Dirac string}
\label{eq7}\end{equation}

However $F^a_{\mu\nu}$, eq.(\ref{eq1}), can also be defined in the Coulomb phase, where monopoles do not exist as solitons,
by assuming $\phi^a$ of the form eq.(\ref{eq2}), transforming in the adjoint representation,
with $U(x)$ an arbitrary gauge transformation. $U(x)$ identifies the abelian projected system, or the 
system in which $\phi^a$ is diagonal: $U(x)$ can be taken as the operator which diagonalizes $\phi$,
the Higgs field, but any other choice is legitimate.

The order parameter for a possible dual superconductiity will be the vev of an operator which creates
a magnetic charge.

Such an operator exists\cite{vari} and has the form
\begin{equation}
\mu^a(\vec x,t) =
e^{
i\int d^3\vec y\,Tr\left(\phi^a(\vec y,t)\vec E(\vec y,t)\right)
\vec b_\perp(\vec x -\vec y)
}
\label{eq8}\end{equation}
\[ \vec\nabla\vec b_\perp = 0\quad, \vec\nabla\wedge \vec b_\perp
= \frac{2\pi}{g}\frac{\vec r}{r^3} +\hbox{Dirac string}\]
$\mu^a$ is gauge invariant and color singlet.

In the abelian projected gauge
$\phi^a = \phi^a_{diag}$, $tr\left\{\phi^a E\right\}  = E^a$
(component of the field along the root $\alpha^a$, and
\begin{equation}
\mu^a(\vec x,t) =
\exp\left\{i\int d^3\vec y\,{\vec E}_\perp^a(\vec y,t)
\vec b_\perp(\vec x -\vec y)\right\}
\label{eq9}\end{equation}
Only ${\vec E}^a_\perp$ survives in the convolution with
$\vec b_\perp$. In any quantization procedure 
${\vec E}^a_\perp$ is the conjugate momentum to ${\vec A}^a_\perp$,
and the operator eq.(\ref{eq9}) is the translation operator
of ${\vec A}^a_\perp$,or, in the Schr\"odinger representation
\[
\mu^a(\vec x,t) | {\vec A}^a_\perp(\vec y,t)\rangle
=  | {\vec A}^a_\perp(\vec y,t) + \vec b_\perp(\vec x - \vec y)\rangle
\]
i.e. $\mu^a(\vec x,t) $ creates a monopole. Of course $\mu^a$ depends on
$\phi^a(x)$, i.e. on the abelian projection $U(x)$.
\section{QCD} 
In QCD there is no Higgs field, but 
$F^a_{\mu\nu}$ can be defined anyhow by eq.(\ref{eq8}), with
$\phi^a = U(x) \phi^a_{diag} U^\dagger(x)$ a scalar field in the adjoint representation
belonging to the orbit of $\phi^a_{diag}$. $U(x)$ identifies the representation in which $\phi^a$ 
is diagonal
(abelian projection), which can be chosen to coincide with the one in which any local
operator $O(x)$ in the adjoint representation is diagonal\cite{hooft81}.
The construction is independent of the properties of $O(x)$ under Lorentz group or discrete 
transformations\cite{kovner}

\begin{figure}
\includegraphics[angle=270,scale=0.4]{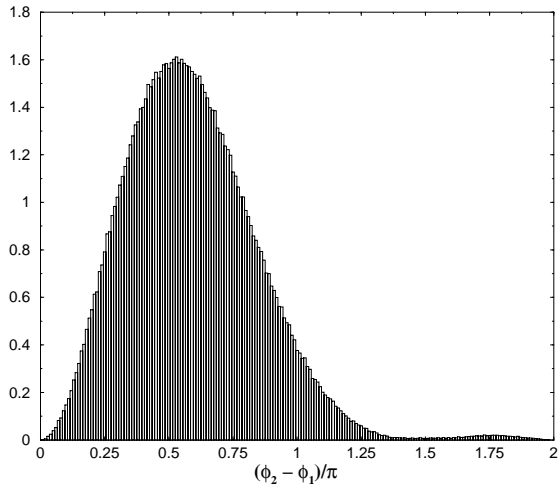}
\caption{An example of probability distribution of the difference of the two highest
eigenvalues of the phase $\Phi$ of the  Polyakov line $e^{i\Phi}$, at the lattice
sites. $SU(3)$ gauge group, $\beta = 6.4$, lattice $16^4$, $10^3$ configurations.}
\end{figure}

The corresponding operator $\mu^a$, eq.(\ref{eq8}), by use of the cyclic invariance of the trace 
can then be written\cite{adg}
\begin{equation}
\mu^a(\vec x,t) =
e^{
i\int d^3\vec y\,Tr\left(\phi^a_{diag}U^\dagger(\vec y,t)\vec E(\vec y,t)
U(\vec y,t)
\right)
\vec b_\perp(\vec x -\vec y)
}
\label{eq10}\end{equation}
If $U(x)$ is independent of the field configuration it can be reabsorbed by changing variables
in the Feynman integral
 by a gauge transformation, with jacobian equal to one, when computing correlators of
$\mu^a$. To all effects
\[ 
\mu^a(\vec x,t) =
e^{
i\int d^3\vec y\,Tr\left(\phi^a_{diag}\vec E_\perp(\vec y,t)
\right)
\vec b_\perp(\vec x -\vec y)
}
\]
and correlations do not depend on $U(x)$.

If $U(x)$ depends  on the gauge fields the jacobian can be non trivial
and the effective lagrangean of the order parameter may depend on it.

However if the bulk density of monopoles is finite the gauge transformation between two different abelian
projections will be continuous eccept at a finite number of points at a given time, i.e. 
at the location of monopoles and will preserve topology. $\mu^a$, eq.(\ref{eq10}) will then create a monopole in all abelian projections. $\langle \mu^a\rangle\neq0$ signals then dual superconductivity in all abelian projections.
This is confirmed by numerical simulations on the lattice, showing that $\langle\mu\rangle$
is independent on the abelian projection\cite{delia}.
The density of monopoles is indeed finite as shown in fig.1 where the distribution of the difference
of eigenvalues of the Polyakov line is displayed. The number of sites where two of them
coincide is zero. This happens for many choices of the operator and different values of the lattice spacing. 
See also re.\cite{mueller}
for the 
density of monopoles in the
maximal abelian gauge.

Dual superconductivity is an intrinsic property independent of the choice of the abelian projection.

We thank M. D'Elia for the measurement reported in fig.1.

This work is partially supported by MIUR project
Teoria delle interazioni fondamentali.

\end{document}